\documentclass[aps,prl,twocolumn,showpacs,superscriptaddress,showkeys,bm,amsmath,amssymb]{revtex4-1}
\usepackage{color}
\usepackage{graphicx}

\begin{document}
\title{Fuel-composition dependent reactor antineutrino yield at RENO}
\affiliation{Institute for Universe and Elementary Particles, Chonnam National University, Gwangju 61186, Korea          }
\affiliation{Institute for High Energy Physics, Dongshin University, Naju 58245, Korea                     }
\affiliation{GIST College, Gwangju Institute of Science and Technology, Gwangju 61005, Korea         }
\affiliation{Institute for Basic Science, Daejeon 34047, Korea     }
\affiliation{Department of Physics, KAIST, Daejeon 34141, Korea          }
\affiliation{Department of Physics, Kyungpook National University, Daegu 41566, Korea          }
\affiliation{Department of Physics and Astronomy, Seoul National University, Seoul 08826, Korea }
\affiliation{Department of Fire Safety, Seoyeong University, Gwangju 61268, Korea              }
\affiliation{Department of Physics, Sungkyunkwan University, Suwon 16419, Korea                }

\author{G. Bak}
\affiliation{Institute for Universe and Elementary Particles, Chonnam National University, Gwangju 61186, Korea          }
\author{J. H. Choi}
\affiliation{Institute for High Energy Physics, Dongshin University, Naju 58245, Korea                     }
\author{H. I. Jang}
\affiliation{Department of Fire Safety, Seoyeong University, Gwangju 61268, Korea              }
\author{J. S. Jang}
\affiliation{GIST College, Gwangju Institute of Science and Technology, Gwangju 61005, Korea         }
\author{S. H. Jeon}
\affiliation{Department of Physics, Sungkyunkwan University, Suwon 16419, Korea                }
\author{K. K. Joo}
\affiliation{Institute for Universe and Elementary Particles, Chonnam National University, Gwangju 61186, Korea          }
\author{K. Ju}
\affiliation{Department of Physics, KAIST, Daejeon 34141, Korea                  }
\author{D. E. Jung}
\affiliation{Department of Physics, Sungkyunkwan University, Suwon 16419, Korea                }
\author{J. G. Kim}
\affiliation{Department of Physics, Sungkyunkwan University, Suwon 16419, Korea                }
\author{J. H. Kim}
\affiliation{Department of Physics, Sungkyunkwan University, Suwon 16419, Korea                }
\author{J. Y. Kim}
\affiliation{Institute for Universe and Elementary Particles, Chonnam National University, Gwangju 61186, Korea          }
\author{S. B. Kim}
\affiliation{Department of Physics and Astronomy, Seoul National University, Seoul 08826, Korea }
\author{S. Y. Kim}
\affiliation{Department of Physics and Astronomy, Seoul National University, Seoul 08826, Korea }
\author{W. Kim}
\affiliation{Department of Physics, Kyungpook National University, Daegu 41566, Korea          }
\author{E. Kwon}
\affiliation{Department of Physics and Astronomy, Seoul National University, Seoul 08826, Korea }
\author{D. H. Lee}
\affiliation{Department of Physics and Astronomy, Seoul National University, Seoul 08826, Korea }
\author{H. G. Lee}
\affiliation{Department of Physics and Astronomy, Seoul National University, Seoul 08826, Korea }
\author{Y. C. Lee}
\affiliation{Department of Physics and Astronomy, Seoul National University, Seoul 08826, Korea }
\author{I. T. Lim}
\affiliation{Institute for Universe and Elementary Particles, Chonnam National University, Gwangju 61186, Korea          }
\author{D. H. Moon}
\affiliation{Institute for Universe and Elementary Particles, Chonnam National University, Gwangju 61186, Korea          }
\author{M. Y. Pac}
\affiliation{Institute for High Energy Physics, Dongshin University, Naju 58245, Korea                     }
\author{Y. S. Park}
\affiliation{Institute for Universe and Elementary Particles, Chonnam National University, Gwangju 61186, Korea          }
\author{C. Rott}
\affiliation{Department of Physics, Sungkyunkwan University, Suwon 16419, Korea                }
\author{H. Seo}
\affiliation{Department of Physics and Astronomy, Seoul National University, Seoul 08826, Korea }
\author{J. W. Seo}
\affiliation{Department of Physics, Sungkyunkwan University, Suwon 16419, Korea                }
\author{S. H. Seo}
\affiliation{Department of Physics and Astronomy, Seoul National University, Seoul 08826, Korea }
\author{C. D. Shin}
\affiliation{Institute for Universe and Elementary Particles, Chonnam National University, Gwangju 61186, Korea          }
\author{J. Y. Yang}
\affiliation{Department of Physics and Astronomy, Seoul National University, Seoul 08826, Korea }
\author{J. Yoo}
\affiliation{Institute for Basic Science, Daejeon 34047, Korea     }
\affiliation{Department of Physics, KAIST, Daejeon 34141, Korea                  }
\author{I. Yu}
\affiliation{Department of Physics, Sungkyunkwan University, Suwon 16419, Korea                }

\collaboration{The RENO Collaboration}
\noaffiliation
\date{\today}

\begin{abstract}
We report a fuel-dependent reactor electron antineutrino ($\overline{\nu}_e$) yield using six 2.8\,GW$_{\text{th}}$ reactors in the Hanbit nuclear power plant complex, Yonggwang, Korea. The analysis uses $850\,666$ $\overline{\nu}_e$ candidate events with a background fraction of 2.0\,\% acquired through inverse beta decay (IBD) interactions in the near detector for 1807.9 live days from August 2011 to February 2018. Based on multiple fuel cycles, we observe a fuel $^{235}$U dependent variation of measured IBD yields with a slope of $(1.51 \pm 0.23) \times 10^{-43}$~cm$^2$/fission and measure a total average IBD yield of $(5.84 \pm 0.13) \times 10^{-43}$~cm$^2$/fission. The hypothesis of no fuel-dependent IBD yield is ruled out at 6.6\,$\sigma$.  The observed IBD yield variation over $^{235}$U isotope fraction does not show significant deviation from the Huber-Mueller (HM) prediction at 1.3\,$\sigma$. The measured fuel-dependent variation determines IBD yields of $(6.15 \pm 0.19) \times 10^{-43}$~cm$^2$/fission and $(4.18\pm 0.26) \times 10^{-43}$~cm$^2$/fission for two dominant fuel isotopes $^{235}$U and $^{239}$Pu, respectively. The measured IBD yield per $^{235}$U fission shows the largest deficit relative to the HM prediction. Reevaluation of the $^{235}$U IBD yield per fission may mostly solve the Reactor Antineutrino Anomaly (RAA) while $^{239}$Pu is not completely ruled out as a possible contributor of the anomaly. We also report a 2.9\,$\sigma$ correlation between the fractional change of the 5\,MeV excess and the reactor fuel isotope fraction of $^{235}$U.
\end{abstract}
\maketitle

A definitive measurement of the smallest neutrino mixing angle $\theta_{13}$ is a tremendous success in neutrino physics during the last decade~\cite{RENO2012, DayaBay2012}. The measurement has been achieved by comparing the observed $\overline{\nu}_e$ fluxes with detectors placed at two different distances from the reactors. As reactor $\overline{\nu}_e$ experiments suffer from large reactor related uncertainties of the expected $\overline{\nu}_e$ flux and energy spectrum~\cite{Hayes2014, Mention2011, Huber2011, mueller2011,Abazajian:2012ys}, identical detector configuration is essential to cancel out the systematic uncertainties. The RAA, $\sim$6\,\% deficit of measured $\overline{\nu}_e$ flux compared to the HM prediction, is an intriguing mystery in current neutrino physics research and needs to be understood~\cite{Seo:2014xei, RENO:2015ksa, Mention2011, Huber2011, mueller2011, Schreckenbach:1985ep, Hahn:1989zr}. There have been numerous attempts to explain this anomaly by incorrect inputs to the fission $\beta$ spectrum conversion, deficiencies in nuclear databases, underestimated uncertainties of reactor $\overline{\nu}_e$ model, and the existence of sterile neutrinos~\cite{Hayes2014, Dwyer:2014eka, Hayes:2015yka, Sonzogni:2016yac, Kopp2013, Dentler:2017tkw, Giunti:2017yid, Gariazzo:2018mwd, Gebre:2017vmm}. Moreover, all of ongoing reactor $\overline{\nu}_e$ experiments have observed a 5\,MeV excess in the IBD prompt spectrum with respect to the expected one~\cite{Seo:2014xei,RENO:2015ksa,An:2015nua,Abe:2015rcp}. This suggests that reactor $\overline{\nu}_e$ model is not complete at all. \\
In commercial nuclear reactor power plants, almost all ($>$\,99\,\%) $\bar{\nu}_e$'s are produced through thousands of $\beta$-decay branches of fission fragments from $^{235}$U, $^{239}$Pu, $^{238}$U, and $^{241}$Pu. The $\overline{\nu}_e$ flux calculation is based on the inversion of spectra of the $\beta$-decay electrons of the thermal fissions which were measured in 1980s at ILL~\cite{Schreckenbach:1985ep, Hahn:1989zr}. The reactor $\overline{\nu}_e$ models using these measurements as inputs have large uncertainties~\cite{Huber2011, mueller2011, Abazajian:2012ys}. Therefore, reevaluation of reactor $\overline{\nu}_e$ model and precise measurements of the neutrino flux and spectrum are essential to understand the RAA. Recently, Daya Bay collaboration reported an observation of correlation between the reactor core fuel evolution and changes in the reactor $\overline{\nu}_e$ flux and energy spectrum~\cite{DayaBay2017b}. The collaboration concluded that the $^{235}$U fuel isotope may be the primary contributor to the RAA. In this Letter, we report an observation of a fuel-dependent variation of the reactor $\overline{\nu}_e$ flux using 1807.9 days of Reactor Experiment for Neutrino Oscillation (RENO) near detector data. We also present a hint of correlation between the 5\,MeV excess and the reactor fuel isotope fraction of $^{235}$U.\\

The Hanbit nuclear power plant complex consists of six reactor cores with total 16.8 GW$_{\text{th}}$ in full operation mode. Two identical detectors are located at 294\,m (near detector) and 1383\,m (far detector) from the reactor array center. The near (far) detector is under 120 (450) meters of water-equivalent rock overburden. The detectors with hydrocarbon liquid scintillator (LS) provide free protons as a target. Coincidence between a prompt positron signal and a delayed signal of gammas from neutron capture by Gadolinium (Gd) provides a distinctive IBD signature. Further details of the RENO detectors and $\overline{\nu}_e$ data analysis are found in Ref.~\cite{RENO:2015ksa}.\\
\begin{figure}[t!]
\centering
\includegraphics[width=0.98\linewidth]{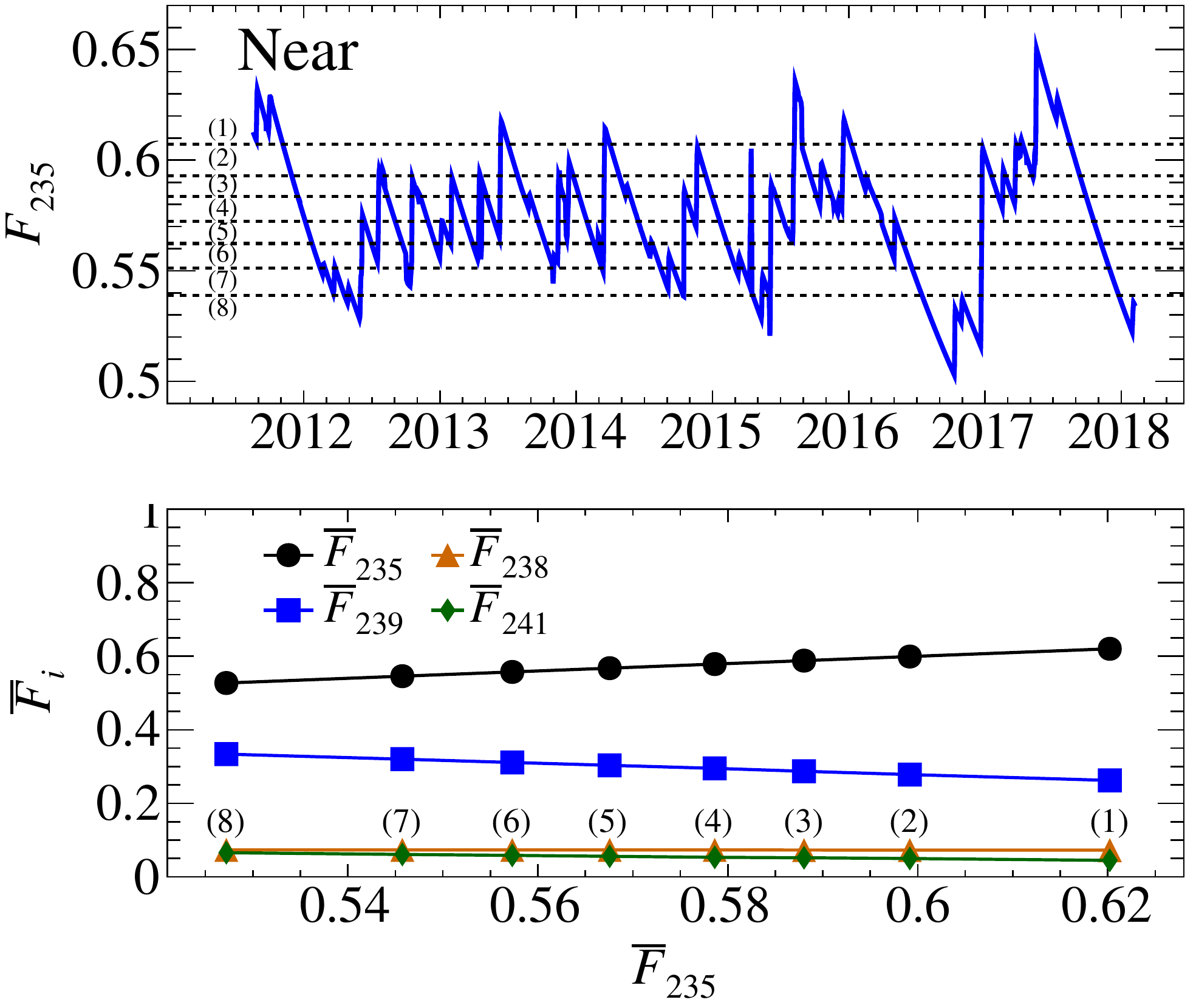}
\caption{Top: Effective $^{235}$U daily fission fraction ($F_{235}$) in the near detector according to Eq.~\eqref{eqn:f235}. The daily $F_{235}$ is obtained from the reactor information provided by the Hanbit nuclear power plant. Bottom: Relative fission fractions for the primary fuel isotopes of $^{235}$U, $^{239}$Pu, $^{238}$U, and $^{241}$Pu. The numbers in the parentheses represent eight data groups with different fission fractions.}
\label{fig:f235}
\end{figure}
The data used in this analysis are taken through IBD interactions in the near detector for 1807.9 live days from August 2011 to February 2018. Applying selection requirements yields $850\,666$ IBD candidates with a background fraction of 2.0\,\%. For the near detector data, we exclude a period of January to December 2013 because of detection inefficiency caused by an electrical noise coming from an uninterruptible power supply. 
 We measure the reactor $\overline{\nu}_e$ flux as a function of an effective fission fraction $F_{i}$(t) given by
\begin{eqnarray}\label{eqn:f235}
F_{i}(t) &= \dfrac{{\sum\limits_{r=1}^{6} \dfrac{W_{{\rm th},r}(t)\overline p_{r}(t)f_{i,r}(t)} {L_{r}^{2}\overline E_{r}(t)}}}{{\sum\limits_{r=1}^{6} \dfrac{W_{{\rm th},r}(t)\overline p_{r}(t)} {L_{r}^{2}\overline E_{r}(t)}}},
\end{eqnarray}

\noindent where $f_{i,r}(t)$ is the fission fraction of $i$-th isotope in the $r$-th reactor, $W_{{\rm th},r}(t)$ is the $r$-th reactor thermal power, $\overline p_{r}(t)$ is the mean survival probability of $\overline{\nu}_e$ from the $r$-th reactor, and $L_{r}$ is the distance between the near detector and the $r$-th reactor. The average $\overline{\nu}_e$ energy produced following fission to be converted into heat is given by $\overline E_{r}(t) = \sum\limits_{i=1}^{4}f_{i,r}(t)\left<E_{i}\right>$ where $\left<E_{i}\right>$ is an average energy released per fission where ($\left<E_{235}\right>$, $\left<E_{238}\right>$, $\left<E_{239}\right>$, $\left<E_{238}\right>$)=(202.36, 205.99, 211.12, 214.26)\, energy per fission/MeV~\cite{Ma:2012bm}. The upper panel of Fig.~\ref{fig:f235} shows time variation of the effective fission fraction of $^{235}$U viewed by the near detector. The effective fission fraction is obtained from the daily thermal power and fission fraction data of each reactor core, provided by the Hanbit nuclear power plant. A total average IBD yield $(\overline{y}_{f})$ is measured to be $\overline{y}_{f} = (5.84\pm 0.13)\times10^{-43}$ cm$^2$/fission with average effective fission fractions $F_{235}$, $F_{238}$, $F_{239}$, and $F_{241}$ of 0.573, 0.073, 0.299, and 0.055, respectively.

For examining fuel-dependent variation of reactor $\overline{\nu}_e$ yield, eight groups of equal data size are sampled according to the eight different values of the $^{235}$U fission fraction. A time-averaged effective fission fraction ($\overline F_{i,j}$) of the $i$-th isotope in the $j$-th data group is calculated as,

\begin{eqnarray}
\overline F_{i,j}&= \dfrac{\int dt \sum\limits_{r=1}^{6} \dfrac{W_{{\rm th},r}(t)\overline p_{r}(t)f_{i,r}(t)} {L_{r}^{2}\overline E_{r}(t)}}{\int dt \sum\limits_{r=1}^{6} \dfrac{W_{{\rm th},r}(t)\overline p_{r}(t)} {L_{r}^{2}\overline E_{r}(t)}}.
\end{eqnarray}\\
The time-averaged effective fission fractions of the four isotopes in each group are shown as a function of time-averaged fission fraction of $^{235}$U ($\overline{F}_{235}$) in the lower panel of Fig.~\ref{fig:f235}. An average IBD yield per fission of the $j$-th data group ($\overline y_{f,j}$) is given by, 
\begin{eqnarray} \label{eqn:Fi}
\overline y_{f,j} = \sum\limits_{i=1}^{4}\overline F_{i,j} \cdot y_{i},
\end{eqnarray}
where an integrated IBD yield per fission ($y_{i}$) is calculated as $y_{i} = \int \sigma(E_{\nu})\phi_{i}(E_{\nu})dE_{\nu}$, $\sigma(E_{\nu})$ is the IBD reaction cross section, and $\phi_{i}(E_{\nu})$ is the reactor $\overline{\nu}_e$ spectrum generated by each reactor's fission isotope, ($y_{235}$, $y_{239}$, $y_{238}$, $y_{241}$)=($6.70\pm0.14$, $4.38\pm0.11$, $10.07\pm 0.82$, $6.07\pm 0.13$)$\times10^{-43}$\,cm$^2$/fission~\cite{Abazajian:2012ys}. We use the IBD cross section in Ref.~\cite{Abazajian:2012ys, Vogel:1999zy} and a neutron lifetime of 880.2\,s in the calculation~\cite{Patrignani:2016xqp}. The IBD yield $y_i$ of a fissile isotope is sensitive to its reactor $\overline{\nu}_e$ spectrum because the IBD cross section increases with the $\overline{\nu}_e$ energy. A model-independent IBD yield of $\overline y_{f,j}$ is determined by counting the number of events in each data group using the following relationship.
\begin{eqnarray} \label{eqn:yi}
N_{j}=\overline y_{f,j} \sum\limits_{r=1}^{6} \dfrac{N_{\rm p}} {4 \pi L^{2}_{r}} \int \left[\dfrac{W_{{\rm th},r}(t)\overline P_{r}(t)} {\sum\limits_{i}f_{i,r}(t)<E_{i}>}\right] \epsilon_{\rm d}(t) dt,
\end{eqnarray}\\
\noindent where $N_{j}$ is the number of IBD events in the $j$-th group, $N_{\rm p}$ is the number of target protons, $\overline P_{r}(t)$ is the mean survival probability, and $\epsilon_{\rm d}(t)$ is the detection efficiency including the signal loss due to timing veto requirements. The average IBD yield of $\overline y_{f,j}$ for each data group is determined by the observed $N_j$. No fission-fraction dependent IBD yield expects a flat distribution of $\overline y_f$ as a function of $\overline{F}_{235}$. There are several updates in this analysis from the previous publication~\cite{Seo:2016uom}. They are use of an IBD cross section in Ref.~\cite{Vogel:1999zy}, an updated detection efficiency including the neutron spill-out effect, and an improved thermal energy release per fission in Ref.~\cite{Ma:2012bm}. A detailed description of the updates will be reported in an upcoming publication.
\begin{figure}[h!]
\centering
\includegraphics[width=0.95\linewidth]{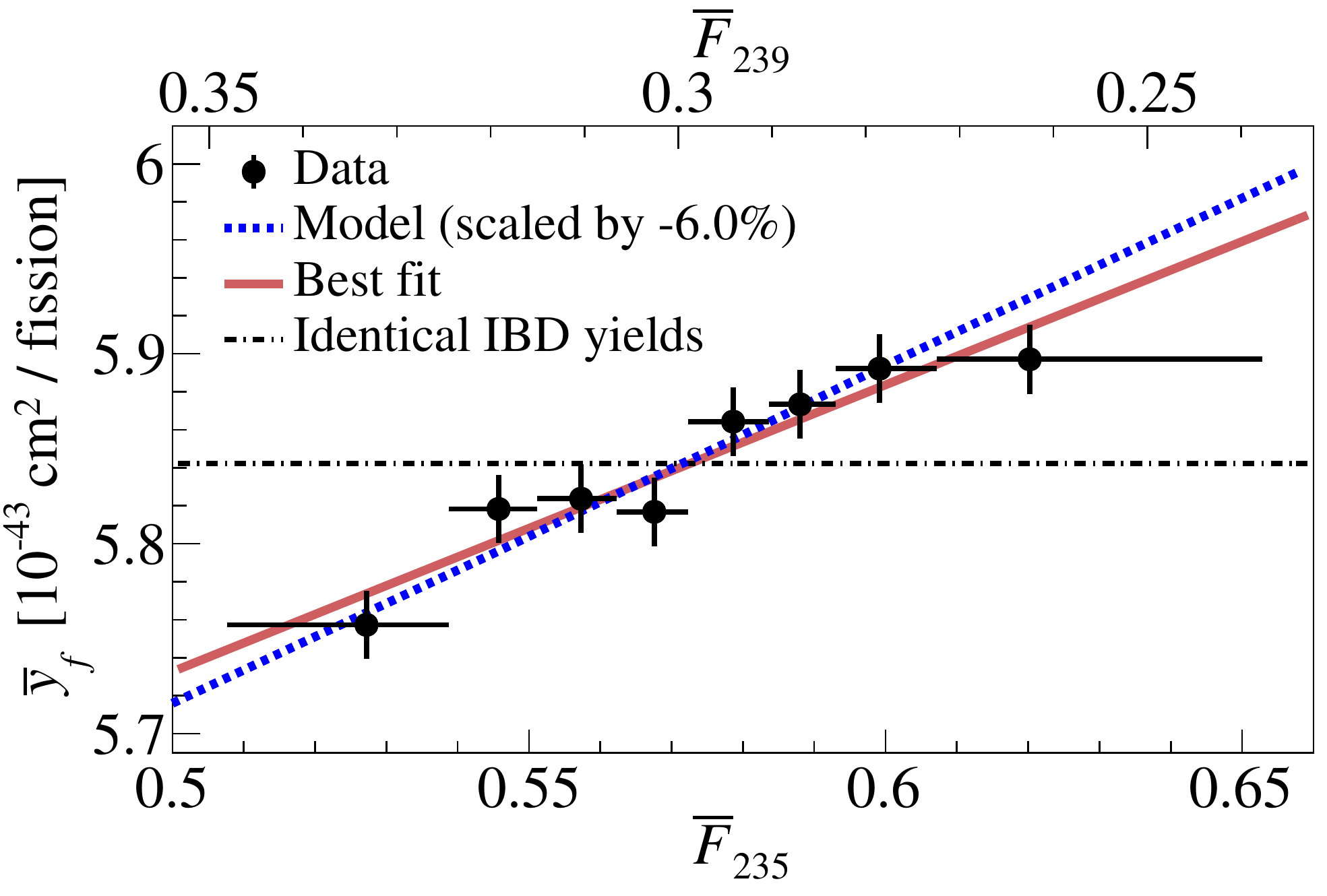}
\caption{IBD yield per fission $\overline{y}_f$ as a function of the $^{235}$U effective fission fraction. The measured values (black dots) are compared to the scaled HM prediction (blue dotted line) and the best fit of the data (red solid line). The value of $\overline{F}_{235}$ for each data point is calculated as an average of fission fractions weighted by thermal power and a distance between reactor and detector. The error of $\overline{F}_{235}$ indicates the variation of $^{235}$U fission fraction. Errors of $\overline{y}_f$ are statistical uncertainties only.}
\label{fig:yi_vs_ff_model}
\end{figure}
Fig.~\ref{fig:yi_vs_ff_model} shows a measured distribution of $\overline y_f$ as a function of $\overline F_{235}$ or $\overline F_{239}$ for the eight data groups. We observe a clear correlation between $\overline y_f$ and $\overline{F}_{235}$, indicating dependence of the IBD yield per fission on the isotope fraction of $^{235}$U. A linear function is used for a fit to the eight data points with $\chi^{2}$/NDF=$4.60/6$ at the best fit. The horizontal line represents an expected distribution for no fuel-dependent IBD yield. This result rules out no fuel-dependent variation of the IBD yield per fission at $6.6\,\sigma$ confidence level, corresponding to the p-value of $3.4 \times 10^{-11}$. It indicates that the variation of the {$\overline y_f$ as a function of ${\overline F}_{235}$ comes from unequal IBD yields among different isotope fissions. The measured yield variation is fitted with the HM prediction to obtain the best-fit at a scaling of -6.0\,\% with $\chi^{2}$/NDF=6.25/7.} Thus the observed IBD yield variation over $^{235}$U fission fraction is not inconsistent with the HM prediction at 1.3\,$\sigma$. The measured IBD yield variation is also fit with the prediction from the ab initio calculation in Ref.~\cite{Hayes:2017res}. A best-fit of $\chi^2$/NDF=4.79/7 is found at a scaling of -5.1\,\% to make a better agreement with the data in the slope.\\

For determination of $y_{235}$ and $y_{239}$ simultaneously, a $\chi^2$ with pull parameter terms of systematic uncertainties is constructed using the observed IBD yield per fission and minimized by varying the free parameters of $y_{235}$ and $y_{239}$, and pull parameters. The subdominant isotopes of $^{238}$U and $^{241}$Pu are constrained in the fitter within uncertainties of 10\,\%~\cite{Mention2011} and 5\,\%~\cite{Vogel:2016ted}, respectively. The uncertainties of thermal power, fission fraction, energy per fission and detection efficiency are considered to be fully correlated among the eight data groups in the different fission fraction bins. Each correlated uncertainty is taken into account through a pull parameter in the $\chi^2$ calculation. The $\chi^2$ is given by
\begin{widetext}
\begin{equation} \label{eqn:chi2}
\begin{aligned}
\chi ^2 = & \sum_{j=1}^{8}\left(  \frac{\overline y_{{\rm obs},j} - \overline y_{{\rm exp},j}} {\sigma_{{\rm obs},j}}  \right)^2 +\left( \frac{\xi_{238}}{\sigma_{238}}  \right) ^2 + \left( \frac{\xi_{241}}{\sigma_{241}}  \right) ^2 + \left( \frac{\xi_{\rm th}}{\sigma_{\rm th}}  \right) ^2 + \left( \frac{\xi_{\rm f}}{\sigma_{\rm f}}  \right) ^2 +\left( \frac{\xi_{\rm en}}{\sigma_{\rm en}}  \right) ^2+ \left( \frac{\xi_{\rm det}}{\sigma_{\rm det}}  \right) ^2\\
&\text{where~~} \overline y_{{\rm exp},j} = \left[\overline F^{j}_{235}\cdot y_{235} + \overline F^{j}_{239}\cdot y_{239} + \overline F^{j}_{238}\cdot y_{238}(1+\xi_{238}) + \overline F^{j}_{241}\cdot y_{241}(1+\xi_{241})\right]\\
&~~~~~~~~~~~~~~~~~~~~~~\cdot (1+\xi_{\rm th}+\xi_{\rm f}+\xi_{\rm en}+\xi_{\rm det}),
\end{aligned}
\end{equation}
\end{widetext}

\begin{figure}[!h]
\centering
\includegraphics[width=0.95\linewidth]{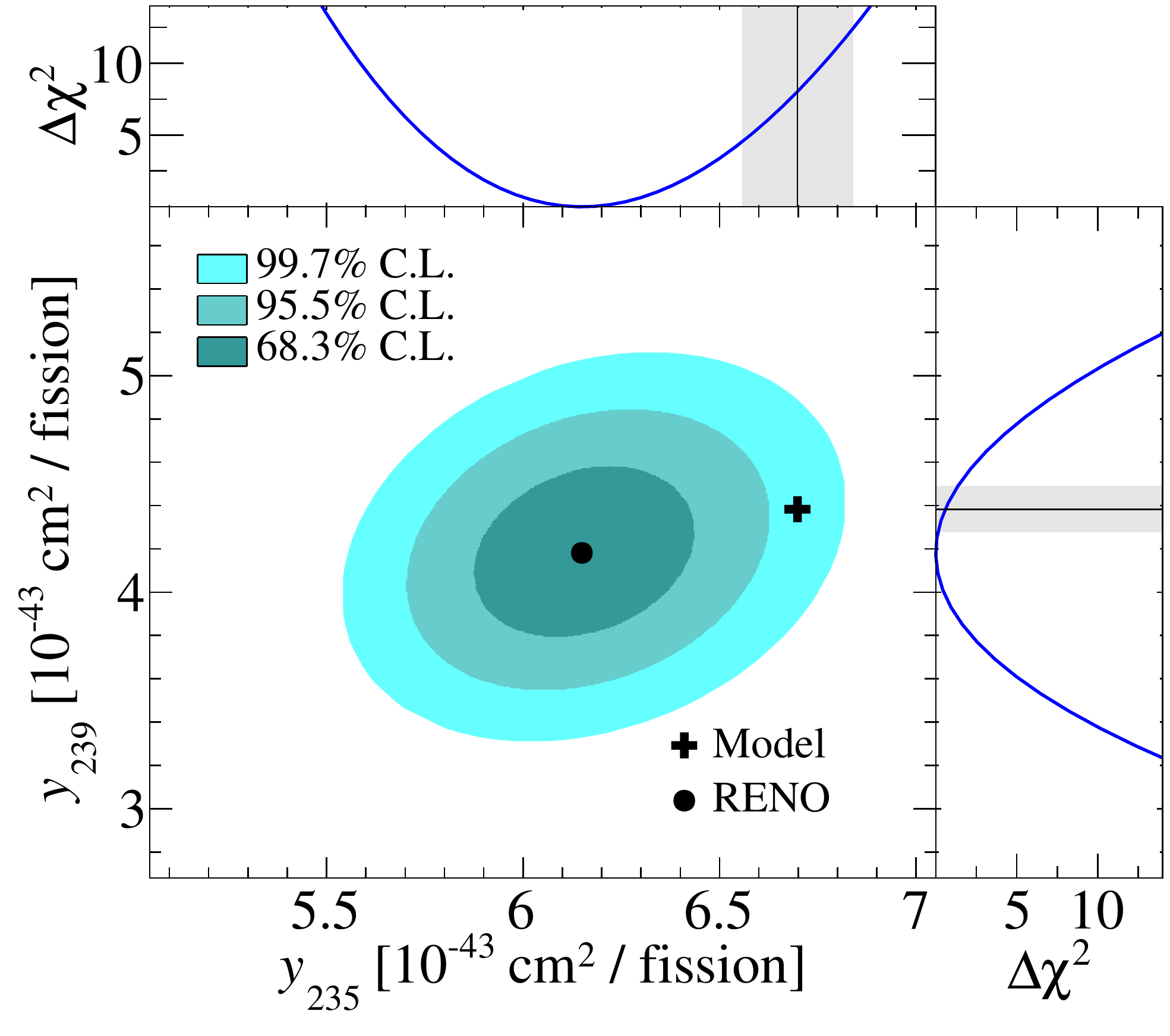}
\caption{Combined measurement of $y_{235}$ and $y_{239}$. The shaded contours are allowed regions and the dot is the best fit. The cross shows the prediction of the HM. The top and right side panels show one dimensional $\Delta \chi^2$ profile distributions for $y_{235}$ and $y_{239}$ while the grey shaded bands represent the model predictions.}
\label{fig:contour}
\end{figure}

\begin{figure*}[t!h]
\centering
\includegraphics[width=0.8\linewidth]{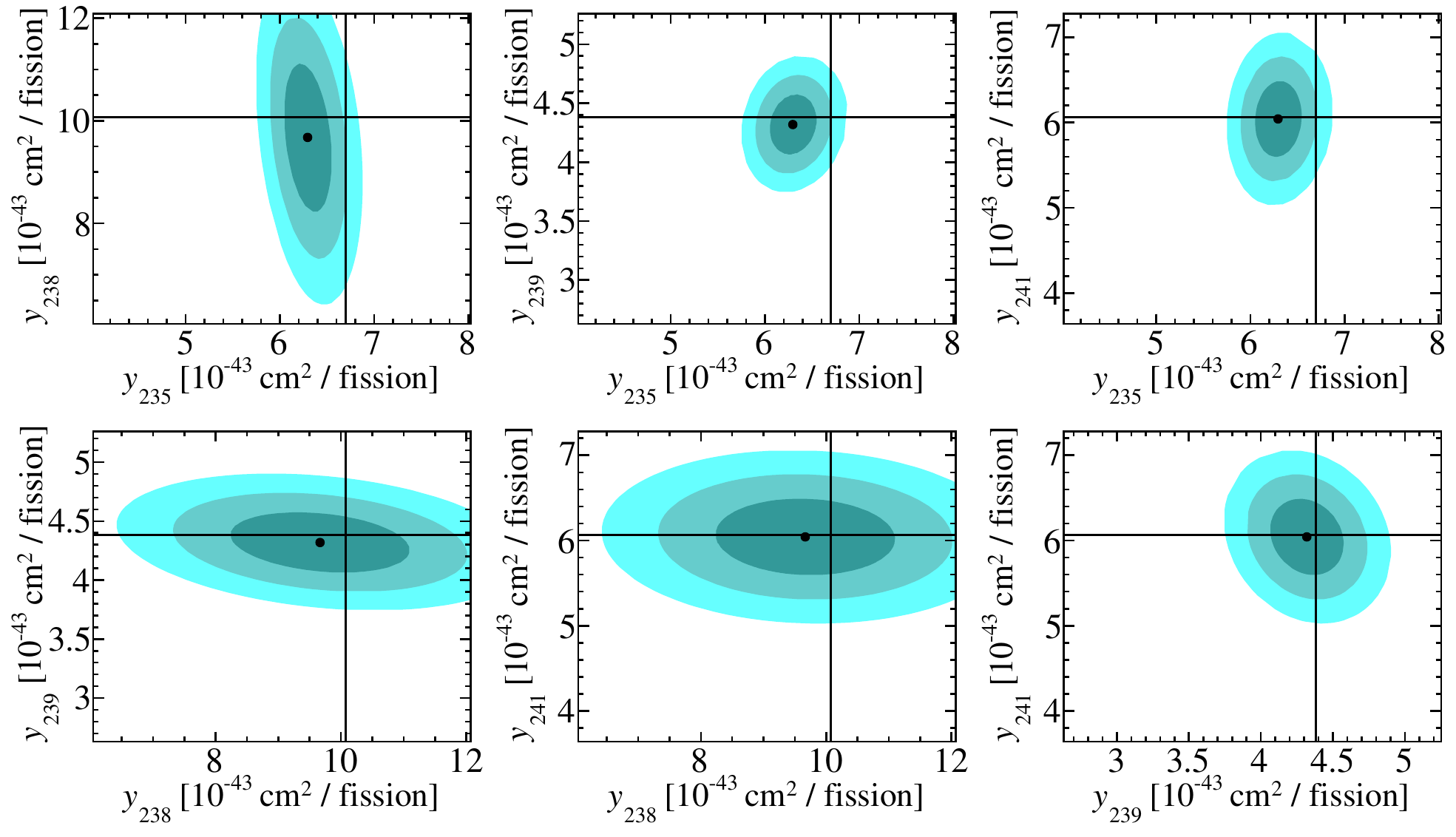}
\caption{Allowed regions of the IBD yield per fission for the six pairs of fission isotopes. The dots indicate the best-fit IBD yields and the cross lines represent the model prediction. The three contours are allowed regions of 68.3, 95.5 and 99.7\,\% C.L.}
\label{fig:contour6}
\end{figure*}

\begin{figure}[h!]
\centering
\includegraphics[width=0.95\linewidth]{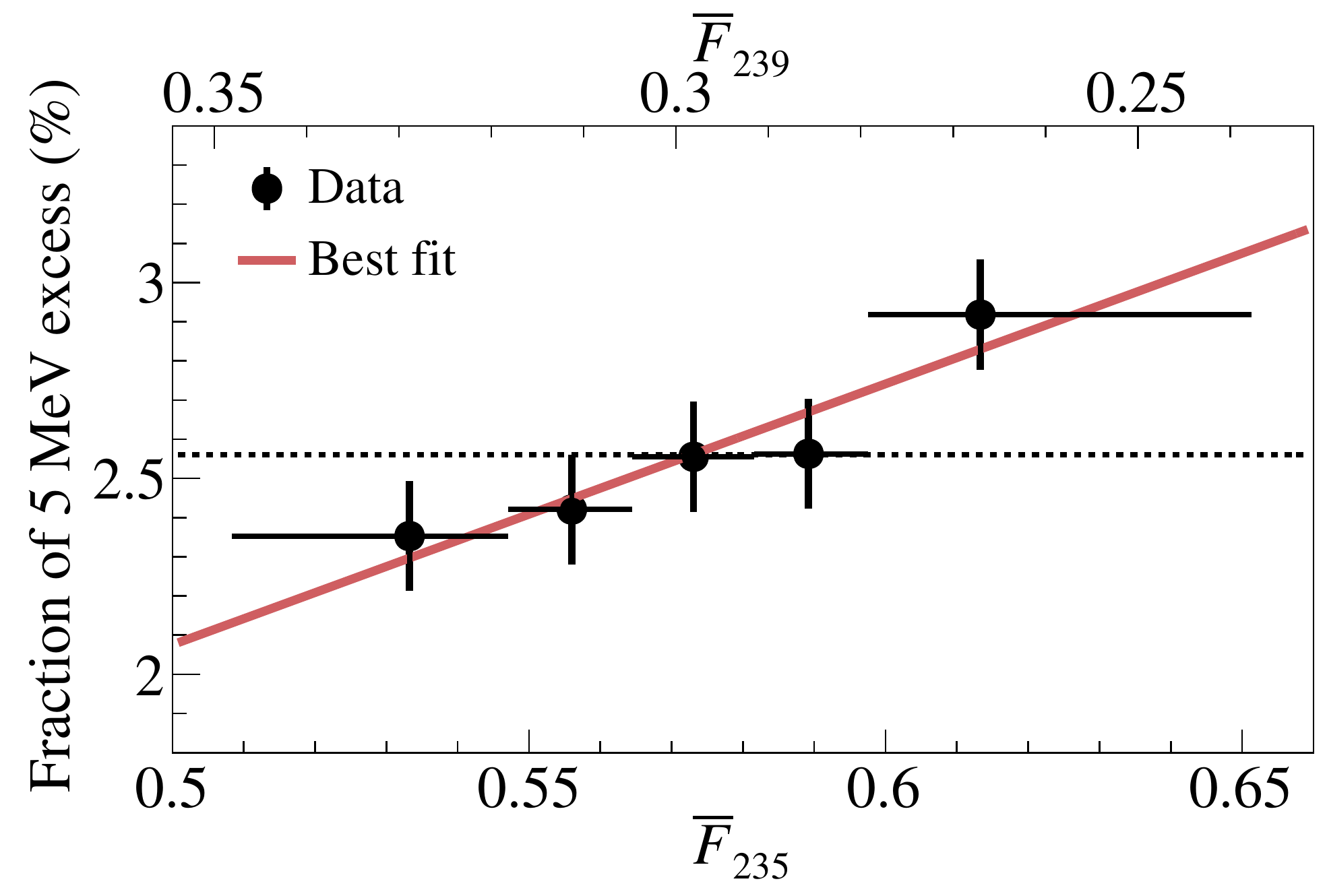}
\caption{Fraction of the 5\,MeV excess as a function of ${\overline F}_{235}$. The red line is the best fit to the data and the dotted line represents no correlation of 5\,MeV excess fraction with ${\overline F}_{235}$.}
\label{fig:5MeV-fueldep-fraction}
\end{figure}

\noindent where $\overline y_{{\rm obs},j}$ is the observed IBD yield per fission averaged over the four isotopes in the $j$-th data group, $\sigma_{{\rm obs},j}$ is the statistical uncertainty of $\overline y_{{\rm obs},j}$, $\overline y_{{\rm exp},j}$ is the expected IBD yield per fission averaged over the four isotopes, $\overline F^{j}_{i}$ is the time-averaged effective fission fraction of the $i$-th isotope for the $j$-th data group, $\sigma_{238}$ and $\sigma_{241}$ are the uncertainties of $y_{238}$ (10\,\%) and $y_{241}$ (5\,\%), respectively, $\sigma_{\rm th}$, $\sigma_{\rm f}$, $\sigma_{\rm en}$ and $\sigma_{\rm det}$ are the uncertainties of thermal power (0.5\,\%), fission fraction (0.7\,\%), energy per fission (0.2\,\%) and detection efficiency (1.93\,\%), respectively. The correlated uncertainties among the eight data groups are considered by changing pull parameters in the $\overline{y}_{exp,j}$ within their uncertainties. Each pull parameter is common among the eight data groups to treat its fully correlated uncertainty. $\xi_{238}$ and $\xi_{241}$ are the pull parameters of $y_{238}$ and $y_{241}$, respectively, and $\xi_{\rm th}$, $\xi_{\rm f}$, $\xi_{\rm en}$ and $\xi_{\rm det}$ are the pull parameters for thermal power, fission fraction, energy per fission and detection efficiency, respectively.

The best-fit results are $y_{235} =(6.15 \pm 0.19) \times 10^{-43}$\,cm$^{2}$/fission and $y_{239} =(4.18 \pm 0.26) \times 10^{-43}$\,cm$^{2}$/fission. Fig.~\ref{fig:contour} shows the combined measurement of $y_{235}$ and $y_{239}$. The measured IBD yield per $^{235}$U fission is smaller than the HM prediction at 2.8\,$\sigma$ while the measured yield per $^{239}$Pu fission is smaller than the prediction only at 0.8\,$\sigma$. This suggests that the RAA can be largely understood by incorrect estimation of the $^{235}$U IBD yield.\\

Following the analysis in Ref.~\cite{Giunti:2016elf} we also perform the combined measurements for all combinations of the four isotopes, total six pairs. The $\chi^2$ of Eq.~\eqref{eqn:chi2} is used with an added constraint term of $({\xi_{i}}/{\sigma_{i}})^2$, where $\sigma_i$ are uncertainties of $y_{235}$ (5\,\%), $y_{238}$ (10\,\%), $y_{239}$ (5\,\%) and $y_{241}$ (5\,\%)~\cite{Giunti:2016elf}. Fig.~\ref{fig:contour6} shows allowed regions of each pair of IBD yields per fission. The dot is the best fit of each pair of IBD yields while the crossing lines represent the HM predicted yields. The shaded contours are 68.3, 95.5 and 99.7\,\% C.L. allowed regions for each pair of IBD yields. In the fitting results of the six pairs of isotopes, we observe that $y_{235}$ is smaller than the prediction at $\sim$2.5\,$\sigma$ while the IBD yields per fission of the rest isotopes are consistent with the prediction within 1\,$\sigma$. \\

The deficit of $y_{235}$ relative to the HM prediction could be interpreted by an indication of incorrectly evaluated IBD yield of $^{235}$U fission that may be a major source of the RAA~\cite{DayaBay2017b, Giunti:2017nww}. As the $^{235}$U and $^{239}$Pu fission fractions are correlated, we perform pseudoexperiments to test this possibility. Pseudodata with IBD yields per fission, ${\overline y}_{f,j}$, are produced for various ratios of $y_{235}$ and $y_{239}$. For each input of $y_{235}$ and $y_{239}$, 1000 pseudodata are produced within statistical errors. In addition, pseudoexperiments with $\overline{y}_f$ scaled down by 6.0\,\% from the model prediction are generated by reducing $y_{235}$ only. A fit finds a value of $y_{235}$ less than the measured value with 3.4\,$\sigma$ deviation from the model prediction. This does not reproduce the measured $y_{235}$ deviation of 2.8\,$\sigma$ by reducing $y_{235}$ only, while a pseudoexperiment of $y_{235}$ down by 8.2\,\% and $y_{239}$ down by 4.6\,\%  reproduce the observed deviations of 2.8\,$\sigma$ and 0.8\,$\sigma$, respectively. Thus, we conclude that RENO data do not rule out $^{239}$Pu as a contributor to the RAA and the anomaly can be mostly explained by reevaluation of the $^{235}$U IBD yield.\\

The RENO collaboration has reported an excess of the observed IBD prompt spectrum at 5\,MeV~\cite{Seo:2014xei,RENO:2015ksa}, also observed by the other ongoing reactor $\overline{\nu}_e$ experiments as well~\cite{An:2015nua,Abe:2015rcp}. The 5\,MeV excess is observed to be proportional to the reactor thermal power~\cite{RENO:2015ksa}. Several explanations and suggestions are proposed to understand the origin of the 5\,MeV excess~\cite{Dwyer:2014eka,Zakari-Issoufou:2015vvp,Hayes:2015yka,Sonzogni:2016yac,Rasco:2016leq}. There is a suggestion that a particular isotope could be the source of the excess~\cite{Hayes:2015yka}, while an analysis disfavors the $^{239}$Pu and $^{241}$Pu isotopes as a single source for the 5\,MeV excess~\cite{Huber:2016xis}. However,  there is no clear understanding of the origin of the 5\,MeV excess yet. \\

A possible fuel dependence of the 5\,MeV excess is examined by the IBD yield per fission for the events only in the 5\,MeV region of $3.8<E_{\rm p}<7$\,MeV. We have not seen any significant deviation of the IBD yield slope with respect to the HM prediction between the 5\,MeV region and the entire energy range of 1.2 to 8\,MeV, consistent with the results from Daya Bay~\cite{DayaBay2017b}. For a fuel-dependence of the 5\,MeV excess only, an event rate with the HM expected energy spectrum in the 5\,MeV region is obtained by a fit to the data in the energy ranges of 1.2 to 3.8\,MeV and 7.0 to 8.0\,MeV and subtracted from the total 5\,MeV rate. Five groups of equal data size are sampled according to five different values of ${\overline F}_{235}$. A fraction of the 5\,MeV excess is calculated as a ratio of the 5\,MeV excess rate to the total IBD rate in the entire energy range. Fig.~\ref{fig:5MeV-fueldep-fraction} shows the distribution of 5\,MeV excess fractions as a function of ${\overline F}_{235}$. The best-fit for the data with a first-order polynomial function shows a correlation between the 5\,MeV excess fraction and ${\overline F}_{235}$. The data is also fitted with a zeroth-order polynomial function with an average excess fraction of $(2.56\pm 0.06)$\,\%. The hypothesis of no-correlation between the 5\,MeV excess fraction and ${\overline F}_{235}$ is disfavored at 2.9\,$\sigma$ where the $\chi^2$/NDF is 1.17/3 for the best fit and 9.58/4 for no-correlation hypothesis. While the current result shows an indicative correlation of the 5\,MeV excess fraction with ${\overline F}_{235}$ and an anti-correlation with the rest isotope fractions, further accumulated data may reveal the source of the 5\,MeV excess. We repeat extraction of the 5\,MeV excess by subtracting the HM prediction estimated from reactor thermal powers and fuel isotope fractions. The significance of correlation between the 5\,MeV excess fraction and $\overline{F}_{235}$ becomes 1.3\,$\sigma$. The data-driven subtraction described earlier is free from the uncertain HM flux normalization.\\

In summary, we report a fuel-dependent IBD yield using 1807.9 days of RENO near detector data. We measure IBD yields per fission of $(6.15\pm 0.19) \times 10^{-43}$~cm$^2$/fission and $(4.18 \pm 0.26) \times 10^{-43}$~cm$^2$/fission for the dominant fission isotopes of $^{235}$U and $^{239}$Pu, respectively. A change in the IBD yield with respect to the effective $^{235}$U fission fraction is observed at 6.6\,$\sigma$. The measured IBD yield per fission of (5.84$\pm$0.13)$\times 10^{-43}$\,cm$^{2}$/fission is 6.0\,\% smaller than the HM prediction and confirms the RAA. The measured IBD yield per $^{235}$U fission is smaller than the HM prediction at 2.8\,$\sigma$. This suggests that the RAA can be largely understood by incorrect estimation of the $^{235}$U IBD yield. We obtain the first hint (2.9\,$\sigma$) for a correlation between the 5\,MeV excess fraction and the $^{235}$U fission fraction.\\

\noindent The RENO experiment is supported by the National Research Foundation of Korea (NRF) grant No. 2009-0083526 funded by the Korea Ministry of Science and ICT. Some of us have been supported by a fund from the BK21 of NRF and Institute for Basic Science grant No.~IBS-R017-G1-2018-a00. We gratefully acknowledge the cooperation of the Hanbit Nuclear Power Site and the Korea Hydro \& Nuclear Power Co., Ltd. (KHNP). We thank KISTI for providing computing and network resources through GSDC, and all the technical and administrative people who greatly helped in making this experiment possible.
\bibliographystyle{h-physrev3}
\bibliography{renofuel}

\end{document}